\documentclass{article}
\usepackage{graphicx}
\usepackage{epsfig}
\usepackage[usenames]{color}
\usepackage{subeqn}
\usepackage{ulem} 

\newcommand{\bracket}[2]{\ensuremath{\langle #1|#2\rangle}}
\newcommand{\grad}{{\rm grad}}
\newcommand{\mdiv}{{\rm div}}
\newcommand{\curl}{{\rm curl}}

\newcommand{\kpar}{{k_\parallel}}

\newcommand{\ff}{{\bf f}}
\newcommand{\bg}{{\bf g}}
\newcommand{\bk}{{\bf k}}
\newcommand{\bn}{{\bf n}}
\newcommand{\br}{{\bf r}}

\newcommand{\bu}{{\bf u}}
\newcommand{\bv}{{\bf v}}
\newcommand{\bx}{{\bf x}}
\newcommand{\by}{{\bf y}}
\newcommand{\bz}{{\bf z}}

\newcommand{\bpi}{\ensuremath{\mbox{\boldmath$\pi$}}}

\newcommand{\cD}{\mathcal{D}}
\newcommand{\cH}{\mathcal{H}}

\title{Quantization of the elastic modes in an isotropic plate}
\author{D. V. Anghel\thanks{National Institute for Physics and Nuclear Engineering--''Horia Hulubei'', Str. Atomistilor no.407, P.O.BOX MG-6, Bucharest - Magurele, Romania} and T. K\"uhn\thanks{Nanoscience Center, Department of Physics, University of Jyv\"askyl\"a, P.O. Box 35, FIN-40014 University of Jyv\"askyl\"a}}

\date{\today}

\begin{document}
%

\maketitle

\begin{abstract}

We quantize the elastic modes in a plate. For this, 
we find a complete, orthogonal set of eigenfunctions of the elastic 
equations and we normalize them. These are the phonon modes in the plate and 
their specific forms and dispersion relations are manifested in low 
temperature experiments in ultra-thin membranes. 
%

\end{abstract}

\section{Introduction} \label{sec_intro}

Nowadays, the devices used in many high-sensitivity applications 
reached such a level of miniaturization, that the wavelength of the 
quantum quasiparticles used in their modeling is comparable to the 
dimensions of the device. The examples we are most familiar with are 
the ultra-sensitive electromagnetic radiation detectors. 
In a very general way of speaking, such detectors consist of 
some thin metallic films, a few tens of nanometers in thickness, 
deposited on a dielectric membrane. The dielectric 
membrane is usually made of SiN$_x$ and has a thickness of the order of 
100 nm. To reach the level of sensitivity and speed required for 
applications, these detectors have to work at sub-Kelvin temperatures 
and at such temperatures the dominant phonon wavelength is comparable 
to the devices' thickness \cite{RevModPhys.78.217.Giazotto,ApplSupercond.5.227.Leivo,ApplPhysLett.72.1305.Leivo,PhysRevLett.81.2958.Anghel,Appl.Phys.Lett.82.293.Anghel,PhysicaB.284.2129.Kuzmin,Luukanen:thesis,Leivo:thesis,Anghel:thesis}. 

To describe the thermal properties of such membranes or detectors, the 
electron-phonon interaction, or in general any interaction of phonons 
with impurities or disorder in the membrane, we have to know the 
phonon modes in the membrane. For this, we have to find the eigenmodes 
of the elastic equations and quantize the elastic field. 

For infinite half-spaces, the quantization has been carried out by 
Ezawa \cite{Ann.Phys.67.438.Ezawa}. 

\subsection{Elastic eigenmodes in plates with parallel surfaces} 

\begin{figure}[t]
\begin{center}
\unitlength1mm\begin{picture}(60,27)(0,0)
\put(0,0){\epsfig{file=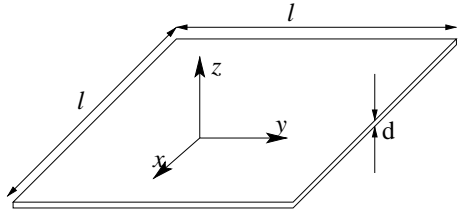,width=60mm}}
\end{picture}
\caption{Typical plate, or membrane, like the ones used to support 
mesoscopic detectors. The thickness $d$ is of the order of 100 nm and 
$l\gg d$. The plate surfaces are parallel to the $xy$ plane and cut the 
$z$ axis at $\pm d/2$.}
\label{plate}
\end{center}
\end{figure}

The elastic eigenmodes in plates with parallel surfaces have been 
studied for a long time, mostly in connection with sound propagation and 
earthquakes (see for example Ref. \cite{Auld:book}). To introduce 
these modes, let us consider a plate of thickness $d$ and area 
$l^2$, with $l\gg d$. The two surfaces of the plate, or membrane, are 
parallel to the $(xy)$ plane and cut the $z$ axis at $\pm d/2$ 
(see Fig. \ref{plate}). Throughout the paper we shall use $V$ for 
the volume of the membrane (or in general of the solid that we describe--see 
Section \ref{selfadjoint}) and $\partial V$ for its surface. We shall 
assume that $l$ is much bigger than any wavelength of the elastic 
perturbations considered here. The 
\textit{displacement field} at \textit{position} $\br$ is 
going to be denoted by  $\bu(\br)$ or $\bv(\br)$. The unit vectors along the 
coordinate axes are denoted by $\hat\bx$, $\hat\by$, and $\hat\bz$. 

The displacement fields obey the dymanic equation 
\begin{equation}
\rho\partial^2 u_i/\partial t^2 = c_{ijkl}\partial_j\partial_k u_l 
\,,\ \forall i=1,2,3. \label{wave1}
\end{equation}
\textit{Here, as everywhere in this paper we shall assume summation over
repeated indices}. 
Assuming that the medium is isotropic (see for example Ref. 
\cite{Auld:book} for the constraints on the tensor $[c]$), Eq. 
(\ref{wave1}) is reduced to $\partial^2u_i/\partial t^2=\partial_j p_{ij}$, 
with $p_{ij}$ defined as in \cite{Ann.Phys.67.438.Ezawa}, 
$p_{ij}=\rho^{-1}c_{ijkl}\partial_k u_l=(c_l^2-2c_t^2)(\partial_k u_k)\delta_{ij}+c_t^2(\partial_i u_j+\partial_j u_i)$, and $c_l$, $c_t$ the 
longitudinal and transversal sound velocities, respectively. 

Introducing the operator $\tilde L$ (we shall use \textit{tilde} for 
operators and \textit{hat} for unit vectors) by 
$\tilde L\bu\equiv \rho^{-1}c_{ijkl}\partial_j\partial_k u_l= c_l^2\grad\,\mdiv\bu-c_t^2\curl\,\curl\bu\equiv c_l^2\nabla\cdot\nabla\cdot\bu-c_t^2\nabla\times\nabla\times\bu$, the wave equation (\ref{wave1}) becomes 
\begin{equation}
\partial^2 \bu/\partial t^2 = \tilde L\bu . \label{wave2}
\end{equation} 
The surface is free, so the stress should be zero there. This amounts to 
the boundary conditions \cite{Auld:book,Ann.Phys.67.438.Ezawa}
\begin{equation}
p_{ij}n_j = 0,\ {\rm on}\ \partial V,\ \forall i=1,2,3, 
\label{boundry_cond}
\end{equation}
where $\hat\bn$ is the unit vector normal to the surface, of components 
$n_1,n_2$, and $n_3$--\textit{we shall use this notation throughout the paper}.

Applying Eqs.\ (\ref{wave2}) and (\ref{boundry_cond}) to the plate 
we obtain the elastic eigenmodes, which are classified 
in three groups, according to their symmetry or polarization direction: 
the \textit{horizontal shear} ($h$), the \textit{symmetric} ($s$) and 
the \textit{antisymmetric} ($a$) waves. 
All these waves are propagating (or decaying, if the wave-vector is complex) 
along the membrane 
and have a stationary form in the direction perpendicular to the surfaces. 
The $h$ wave is polarized parallel to the 
surfaces and perpendicular to the propagation direction. The $s$ and 
$a$ waves are superpositions of longitudinal and transversal waves, 
polarized in a plane that is perpendicular to the surfaces and contains the 
propagation direction. The difference between the $s$ and the $a$ waves 
comes from the fact that the displacement field along the $z$ direction is 
symmetric for the $s$ wave and antisymmetric for the $a$ wave, while 
the displacement field along the propagation direction is antisymmetric 
for the $s$ wave and symmetric for the $a$ wave (see below). 
Explicitely, the three types of modes are: 
\begin{subequations} \label{eigenmodes}
\begin{eqnarray}
\bu_{h}&=& (\hat\bk_\parallel\times\hat\bz)\cdot N_{h} 
\cos\left[\frac{m\pi}{b}\left(z-\frac{b}{2}\right)\right] 
e^{i(\bk_\parallel\cdot\br-\omega t)},\quad m=0,1,2,\ldots, \label{vxsh} \\
\bu_{s} &=& N_s\left\{\hat\bz\cdot k_\parallel\left[-2k_t k_l 
\cos\left(k_t\frac{b}{2}\right) 
\sin\left(k_l z\right) + [k_t^2-k_\parallel^2]\cos\left(k_l 
\frac{b}{2}\right)\sin\left(k_t z\right)\right] 
\right. \nonumber \\
&& \left.+ \hat\bk_\parallel\cdot ik_t\left[2 k_\parallel^2
\cos\left(k_t\frac{b}{2}\right) 
\cos\left(k_l z\right) + [k_t^2-k_\parallel^2]\cos\left(k_l 
\frac{b}{2}\right)\cos\left(k_t z\right)\right]\right\} \nonumber \\
&& \times e^{i (\bk_\parallel\cdot\br-\omega t)} \,, \label{vsym0} \\
\bu_{a}&=&N_a\left\{\hat\bz\cdot k_\parallel\left[2k_tk_l\sin(k_t
\frac{b}{2})\cos(k_lz)-[k_t^2-k_\parallel^2]\sin(k_l\frac{b}{2})
\cos(k_tz)\right] \right. \nonumber \\
&& \left.+\hat\bk_\parallel\cdot ik_t\left[2k_\parallel^2\sin(k_t\frac{b}{2})
\sin(k_lz) 
+[k_t^2-k_\parallel^2]\sin(k_l\frac{b}{2})\sin(k_tz)\right]\right\}
\nonumber \\
&& \times e^{i(\bk_\parallel\cdot\br-\omega t)} , \label{vasym0}
\end{eqnarray}
\end{subequations}
where $\hat\bk_\parallel$ is the unit vector along the propagation direction 
and $\bk_\parallel=k_\parallel\cdot\hat\bk_\parallel$. 
In the $s$ and $a$ modes, the wavevector in the $z$ direction takes two 
values, $k_t$ and $k_l$, one corresponding to the transversal component of 
the mode, the other one corresponding to the longitudinal component. 
The constants $N_{h}$, $N_{s}$, and $N_{a}$ are the normalization constants, 
which will be calculated in Section \ref{normalization}. 

The components of the wavevectors, $k_t$, $k_l$, and $k_\parallel$ 
obey the transcendental equations \cite{Auld:book}, 
\begin{subequations} \label{transcendental}
\begin{eqnarray}
\frac{\tan(\frac{d}{2}k_t)}{\tan(\frac{d}{2}k_l)}&=&-\frac{4k_lk_tk_\parallel^2}{[k_t^2-k_\parallel^2]^2}\label{eqn_symmetric}
\end{eqnarray}
and 
\begin{eqnarray}
\frac{\tan(\frac{d}{2}k_t)}{\tan(\frac{d}{2}k_l)}&=&-\frac{[k_t^2-k_\parallel^2]^2}{4k_lk_tk_\parallel^2} .\label{eqn_antisymmetric}
\end{eqnarray}
\end{subequations}
for symmetric and antisymmetric waves, respectively. 

All the solutions of the elastic equation (\ref{wave2}) are given by 
equations (\ref{vxsh}), (\ref{vsym0}), and (\ref{vasym0}), with $m$ taking all 
the natural values, $0,1,\ldots$, whereas $k_t$ and $k_l$ 
are the solutions of Eqs.\ (\ref{eqn_symmetric}) and 
(\ref{eqn_antisymmetric}). 
$k_\parallel$ can be a complex number, 
but the complete, orthogonal set of phonon modes that we shall use
in the quantization of the elastic field are the ones with
$k_\parallel$ running from $0$ to $\infty$

The paper is organized as follows: 
in section \ref{selfadjoint} we show that $\tilde L$ is self-adjoint 
even when applied to the displacement field of an elastic body of 
arbitrary shape. Therefore, we 
can form a complete, orthonormal set of its eigenfunctions. 

The fact that the elastic modes (\ref{eigenmodes}) corresponding to 
different quantum numbers are orthogonal to each other is proved in 
Section \ref{orthogonality}, based on the hermiticity of $\tilde L$. 

The normalization constants are calculated in Section \ref{normalization} 
and the formal procedure of quantising the elastic field is presented in 
Section \ref{quantization}. 

We apply this formalism elsewhere to 
calculate the phonon scattering in amorphous, thin membranes.

\section{Orthogonality and completeness of the set of elastic eigenmodes}
\label{COS}

\subsection{Self-adjointness of the operator $\tilde L$} \label{selfadjoint}

We shall prove the self-adjointness of $\tilde L$ on an arbitrary volume $V$. 
We assume that $V$ has the smooth border, $\partial V$. The operator 
$\tilde L$ acts on the Hilbert space $\cH$ which consists of the 
vector functions defined on $V$, which are integrable in modulus square. 
The scalar product on $\cH$ is defined as usual, 
\begin{equation}
\langle\bv|\bu\rangle = \int_V\bv^\dagger(\br)\cdot\bu(\br)\, d^3\br \,,
\label{scalar_prod} 
\end{equation}
and the norm is $||\bu||\equiv\langle\bu|\bu\rangle^{1/2}$. 
The domain of $\tilde L$, denoted by $\cD(\tilde L)$, is formed of such 
functions $\bu(\br)\in\cH$, so that $\tilde L\bu(\br)$ exists and is 
contained in $\cH$. Moreover, the functions from 
$\cD(\tilde L)$ should obey the boundary conditions (\ref{boundry_cond}). 
Ezawa showed that $\tilde L$ is hermitian \cite{Ann.Phys.67.438.Ezawa}. 
We will show now that $L$ is also self-adjoint. 

First of all, for a formal treatment, in this section we will understand 
the derivatives in a generalized sense. 
If $f(\br)$ is an arbitrary function on $V$, then 
$\partial_i f$ is defined as the function that satisfies
\begin{equation}
\int_V g(\br)\partial_i f(\br)\,d^3\br = 
\int_{\partial V} f(\br)g(\br) n_i\, d^2\br - 
\int_V f(\br)\partial_i g(\br)\,d^3\br \,, \label{generalderiv}
\end{equation}
for \textit{any} function $g(\br)$ of class $C^1(V)$ (i.e. $g(\br)$ is 
derivable, with continous first derivatives on $V$) and all the integrals 
on the right-hand side of Eq. (\ref{generalderiv}) exist and are finite. 

Returning to our operator, let us first note that the space 
$\cD(\tilde L)$ includes the space of functions twice derivable, with 
continous, integrable, second derivatives, $C^2(V)$, which is dense in 
$\cH$. Therefore 
$\cD(\tilde L)$ is also dense in $\cH$, and $\tilde L$ is a 
\textit{symmetric operator}. 
Then we define the 
adjoint operator $\tilde L^\dagger$ and its domain $\cD(\tilde L^\dagger)$. 
For this, let $\bv$ be a function in $\cH$, so that 
$\bracket{\bv}{\tilde L\bu}$, is a continous linear functional in 
$\bu\in\cD(\tilde L)$, i.e. there exists an $M_\bv>0$ for which 
\begin{equation}
\bracket{\bv}{\tilde L\bu}\le M_\bv||\bu|| \label{defdomadj}
\end{equation}
for any $\bu\in\cD(\tilde L)$. 
By the Riesz-Fr\'echet theorem \cite{RieszNagy:book}, there exists a $\bv^*\in\cH$ so that 
$\bracket{\bv}{\tilde L\bu}=\bracket{\bv^*}{\bu}$ for any 
$\bu\in\cD(\tilde L)$. The adjoint operator $\tilde L^\dagger$ 
is defined by the relation $\tilde L^\dagger\bv=\bv^*$, for all 
$\bv$ that satisfy (\ref{defdomadj}). The functions that satisfy 
(\ref{defdomadj}) form the domain of $\tilde L^\dagger$, denoted 
$\cD(\tilde L^\dagger)$. Since $\tilde L$ is hermitian, 
if $\bv,\bu\in\cD(\tilde L)$, then 
$\bracket{\bv}{\tilde L\bu}=\bracket{\tilde L\bv}{\bu}$ \textit{is} 
a linear functional, so any function from $\cD(\tilde L)$ is 
included in $\cD(\tilde L^\dagger)$. Therefore we can write  
$\cD(\tilde L)\subset\cD(\tilde L^\dagger)$. 
To prove that 
$\tilde L$ is self-adjoint, we have to show that also 
$\cD(\tilde L^\dagger)\subset\cD(\tilde L)$, so in the end 
$\cD(\tilde L^\dagger)=\cD(\tilde L)$ and $\tilde L=\tilde L^\dagger$. 

For this, let us take 
$\bv\in\cD(\tilde L^\dagger)$ and $\bu\in\cD(\tilde L)$. Integrating by 
parts we get 
\begin{eqnarray}
\rho\langle \bv| \tilde L\bu\rangle
 &=& \int_V v_i^\star c_{ijkl}
\partial_j\partial_k u_l d^3\br\\
 &=&- \int_{\partial V}(\partial_j v_i^\star) c_{ijkl}n_k u_l d^2\br + 
\int_V (\partial_k\partial_j v_i^\star) c_{ijkl} u_l d^3\br \nonumber \\
 &=& - \int_{\partial V}(\partial_j v_i^\star) c_{ijkl}n_k u_l d^2\br + 
\rho\langle \tilde L\bv|\bu\rangle \equiv 
\rho\langle \tilde L^\dagger\bv|\bu\rangle \,, \label{Lhermite}
\end{eqnarray}
where we used $c_{ijkl}n_j\partial_k u_l=0$ on $\partial V$ (Eq. 
\ref{boundry_cond}), and the simplified notation 
\begin{equation}
\rho\langle \tilde L\bv|\bu\rangle\equiv 
\int_V (c_{ijkl}\partial_j\partial_k v_i^\star) u_l d^3\br =
\int_V (c_{lkji}\partial_k\partial_j v_i^\star) u_l d^3\br \,,
\label{Lswitch}
\end{equation}
although $\bv$ is not necesarily a function in $\cD(\tilde L)$. 
The last equality in Eq. (\ref{Lswitch}) is obtained using 
$c_{ijkl}=c_{jikl}=c_{ijlk}=c_{klij}$ \cite{Auld:book} and permuting the 
partial derivatives. 
But Eq. (\ref{Lhermite}) means that 
\begin{equation}
\langle (\tilde L-\tilde L^\dagger)\bv|\bu\rangle = \rho^{-1}
\int_{\partial V}\partial_j v_i^\star n_k c_{ijkl} u_l d^2\br 
\label{Lhermite1}
\end{equation}
for any $\bu\in\cD(\tilde L)$. If $(\partial_j v_i^\star) n_k c_{ijkl}$ is not 
identically zero on $\partial V$, then we can find $\bu$ so that the 
surface integral in (\ref{Lhermite1}) is 
different from zero. This implies that 
$\langle (\tilde L-\tilde L^\dagger)\bv|\bu\rangle\ne 0$, so $\bv\ne 0$ 
on a set of measure larger than zero in the interior of 
$V$, denoted as $V^\circ$. In such a case, we can find a nonempty compact 
set $S\subset V^\circ$ and a function, $\bu'$, which is twice derivable, 
zero outside $S$, and satisfies 
\begin{equation}
\langle(\tilde L-\tilde L^\dagger)\bv|\bu'\rangle\ne 0. \label{uprime}
\end{equation}
Since $\bu'$ is zero outside $S$ and $S$ is a compact set in $V^\circ$, 
this means that both, $\bu'$ and any of its derivatives are zero on 
$\partial V$; therefore $\bu'\in\cD(\tilde L)$. Now, 
$\bu'(\br\in\partial V)=0$, implies that 
\begin{equation}
\int_{\partial V}\partial_j v_i^\star n_k c_{ijkl} (u_l+u'_l) d^2\br = 
\int_{\partial V}\partial_j v_i^\star n_k c_{ijkl} u_l d^2\br 
\label{boundaryuprime}
\end{equation}
whereas (\ref{uprime}) implies that 
\begin{equation}
\langle(\tilde L-\tilde L^\dagger)\bv|(\bu+\bu')\rangle\ne 
\langle(\tilde L-\tilde L^\dagger)\bv|\bu\rangle. \label{uprimeu}
\end{equation}
Since Eq. (\ref{Lhermite1}) should be valid for any function in 
$\cD(\tilde L)$, including $\bu$ and $\bu+\bu'$ and Eq. 
(\ref{boundaryuprime}) is true 
by construction, this implies that (\ref{uprimeu}) is a contradiction; 
therefore $(\partial_j v_i^\star) n_k c_{ijkl}=0$ on $\partial V$. 
Moreover, by the definition of $\cD(\tilde L^\dagger)$ and the Riesz-Fr\'echet 
theorem, $c_{lkji}\partial_k\partial_j v_i\equiv \rho\tilde L^\dagger \bv\in\cH$. 
Therefore $\bv\in\cD(\tilde L)$, so $\cD(\tilde L^\dagger)=\cD(\tilde L)$ 
and $\tilde L=\tilde L^\dagger$, i.e. 
the operator $\tilde L$ is self-adjoint on an arbitrary volume, $V$. 

Going back to the plate with infinite lateral extension, we have 
to find from the wave functions of the form (\ref{eigenmodes}) a 
complete, orthonormal set. We can do that by using the operator 
$\tilde\bk_\parallel\equiv i(\partial_x+\partial_y)$, which is also 
self-adjoint when acting on wave-functions defined on a plate with 
infinite lateral extension, or on a finite, rectangular plate, with 
periodic boundary conditions at the 
edges. Since $\tilde L$ and $\tilde\bk_\parallel$ 
commute and they are both self-adjoint operators, if we find 
\textit{all} the eigenfunctions common to $\tilde L$ and 
$\tilde\bk_\parallel$, then we have a complete set. 
But these functions are given by Eqs.\ 
(\ref{eigenmodes}), with real $k_\parallel$, and $k_t$ and $k_l$ 
satisfying (\ref{transcendental}).

\subsection{Orthogonality of the elastic eigenmodes} \label{orthogonality}

Now we study the orthogonality of the elastic eigenmodes of the plate. 
For this, we write the the functions that appear in Eqs.\ 
(\ref{vxsh})-(\ref{vasym0}) in general as $\bu_{\bk_\parallel,k_t,\sigma}(\br)\equiv \bu_{k_t,\sigma}(z)e^{i\bk_\parallel\cdot\br}$, 
where we separated the $x$ and $y$ dependence 
of the fields from the $z$ dependence and we disregarded the time 
dependence. By $\sigma$ we denote the ``polarization'' $h$, $s$, or $a$. 
We shall not use $k_l$ explicitely in the notations below, since it 
is determined implicitely by $k_t$, $\kpar$, and Eqs.\ 
(\ref{transcendental}). 
The operator $\tilde\bk_\parallel$ has eigenvectors of the form 
$\bg(z)e^{i\bk_\parallel\cdot\br}$, where 
$\bg(z)$ is an arbitrary function of $z$ while the eigenvalue 
$\bk_\parallel$, perpendicular on $z$, has real components along the 
$x$ and $y$ directions. We analyse the common set of 
eigenvectors of $\tilde L$ and $\tilde\bk_\parallel$, so in what follows 
we shall consider only the cases with real wave-vectors $\bk_\parallel$. 
First we observe that 
\begin{equation}
\langle\bu_{\bk_\parallel,k_t,\sigma}|\bu
_{\bk'_\parallel,k_t',\sigma'}\rangle = 2\pi
\delta(\bk_\parallel-\bk'_\parallel)\int_{-b/2}^{b/2} 
\bu^\dagger_{k_t,\sigma}(z)\bu_{k_t',\sigma'} 
(z)\,dz, \label{zintegral}
\end{equation}
so we are left to study the orthogonality of functions with the same 
$\bk_\parallel$. For simplicity we choose 
$\bk_\parallel=\hat\bx\cdot k_\parallel$, so the $h$ waves are polarized 
in the $\hat\by$ direction and the $s$ and $a$ waves have displacement 
fields in the $(xz)$ plane. Since the displacement field of the $h$ waves 
are perpendicular to the displacement fields of the $s$ and $a$ waves, 
any $h$ wave is orthogonal to any $s$ or $a$ wave. Similarly, for 
the same $\bk_\parallel$, the displacement fields of any of the $s$ and 
$a$ waves, although in the same plane, are orthogonal to each-other due 
to their opposite symmetries. We conclude that 
\[
\int_{-b/2}^{b/2}\bu^\dagger_{k_t,\sigma}(z)\bu_{k_t',\sigma'}(z)
\,dz = 0\,, 
\]
for any $k_t$ and $k_t'$, if $\sigma\ne\sigma'$. 
Therefore $\bracket{\bu_{\bk_\parallel,k_t,\sigma}}{\bu_{\bk'_\parallel,k_t',\sigma'}}\propto\delta(\bk_\parallel-\bk'_\parallel)\delta_{\sigma,\sigma'}$. 

We are left to show that $\bracket{\bu_{\bk_\parallel,k_t,\sigma}}{\bu_{\bk_\parallel,k_t',\sigma}}\propto\delta_{k_t,k_t'}$ which follows simply from the hermiticity of the operator $\tilde L$. We calculate the matrix element 
\begin{eqnarray*}
\bracket{\bu_{\bk_\parallel,k_t,\sigma}}{\tilde L\bu_{\bk_\parallel, 
k_t',\sigma}} &=& -\omega^2_{\bk_\parallel,k_t',\sigma}
\bracket{\bu_{\bk_\parallel,k_t,\sigma}}{\bu_{\bk_\parallel, 
k_t',\sigma}} \nonumber \\
&=& \bracket{\tilde L\bu_{\bk_\parallel,k_t,\sigma}}{\bu_
{\bk_\parallel, k_t',\sigma}} = -\omega^2_{\bk_\parallel,k_t,
\sigma} \bracket{\bu_{\bk_\parallel,k_t,\sigma}}{\bu_{\bk_\parallel, k_t',\sigma}}
\end{eqnarray*}
But since for given $\bk_\parallel$ and $\sigma$, the eigenstates of 
$\tilde L$ are not degenerate,
$\bracket{\bu_{\bk_\parallel,k_t,\sigma}}{\bu_{\bk_\parallel, k_t',\sigma}}=0$, unless $k_t=k_t'$ and, of course, $k_l=k_l'$. 
This completes the proof and 
\begin{equation}
\bracket{\bu_{\bk_\parallel,k_t,\sigma}}{\bu_
{\bk'_\parallel, k_t',\sigma'}}\propto\delta_{\sigma,\sigma'}
\delta(\bk_\parallel-\bk'_\parallel)\delta_{k_t,k_t'}
 \,. \label{orthogonalproduct}
\end{equation}

In the next section we calculate the normalization constants. 

\section{Normalization of the elastic modes} \label{normalization}

From this point on we shall assume that the plate has finite 
lateral extensions of area $A=l\times l$ and the eigenfunctions obey 
periodic boundary conditions in the $(xy)$ plane. The volume of the plate is 
$V=Ad$. As a consequence, the values of $\bk_\parallel$ are discrete  
and the scalar product (\ref{orthogonalproduct}) should give 
\begin{equation}
\bracket{\bu_{\bk_\parallel,k_t,\sigma}}{\bu_{\bk'_\parallel,k_t',
\sigma'}}= \delta_{\sigma,\sigma'}\delta_{\bk_\parallel,\bk'_\parallel}
\delta_{k_t,k_t'} \,. \label{orthogonalproductd}
\end{equation}

For horizontal shear waves, the normalization constant is simple to 
calculate. In this case, let us write $\bu_{\bk_\parallel,k_t,h}$ 
simply as $\bu_{\bk_\parallel,m,h}$ (see \ref{vxsh}) and we obtain 
\begin{subequations} \label{Nshset}
\begin{eqnarray}
||\bu_{\bk_\parallel,m,h}||^2 &=& 
(N_{\bk_\parallel,m,h})^2 \frac{V}{2}, \qquad {\rm for}\ 
m>0, \label{normshmge1} \\
&=& (N_{\bk_\parallel,m,h})^2 V, \qquad {\rm for}\ 
m=0. \label{normshmge2} 
\end{eqnarray}
\end{subequations}
So $N_{\bk_\parallel,0,h}=V^{-1/2}$ and 
$N_{\bk_\parallel,m>0,h}=(2/V)^{1/2}$. 

\begin{figure}[t]
\begin{center}
\unitlength1mm\begin{picture}(140,55)(0,0)
\put(0,0){\psfig{file=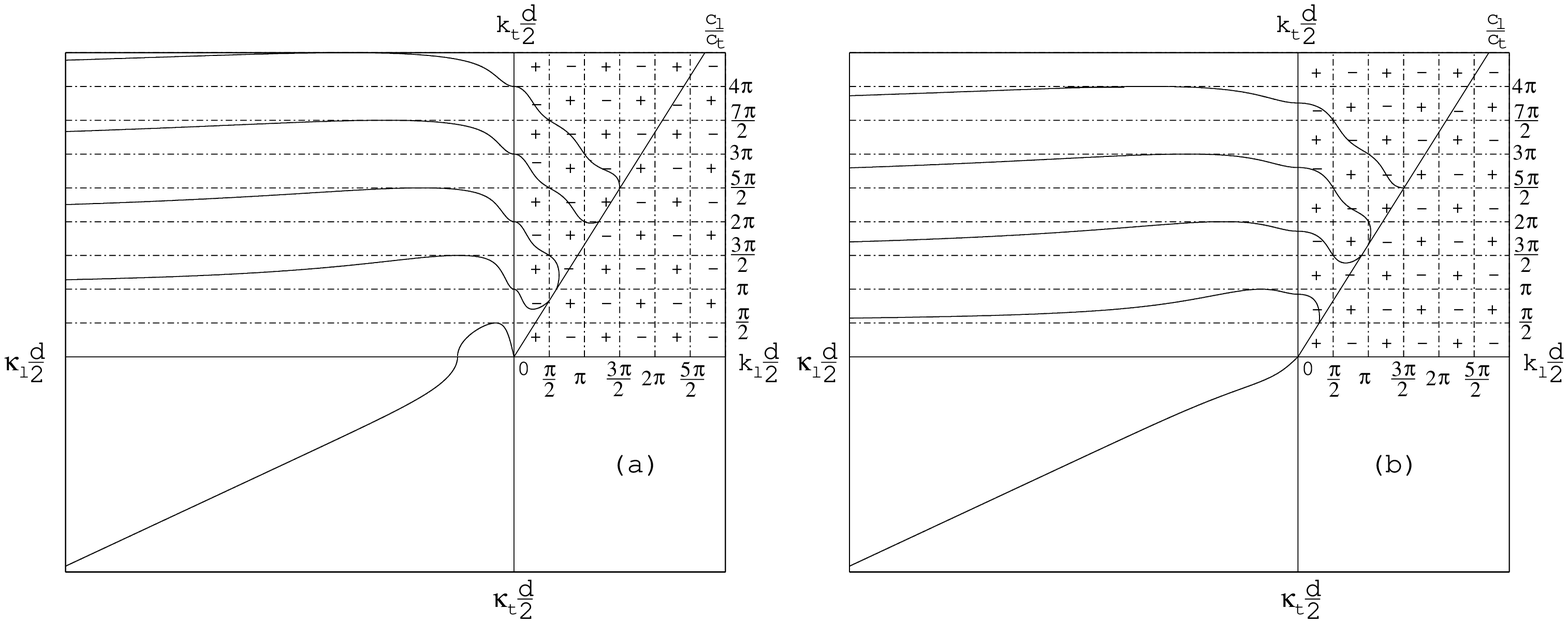,width=120mm}}
\end{picture}
\caption{The curves $(k_l(k_\parallel),k_t(k_\parallel))$ for 
symmetric (a) and antisymmetric (b) Lamb modes. On each curve 
$k_\parallel$ varies from 0 to $\infty$, as we go from its right end to the 
left. The different curves 
correspond to different branches of the dispersion relation, i.e. to 
different values of $k_l$ and $k_t$ at $k_\parallel=0$. Both, 
$k_l$ and $k_t$ may take real and imaginary values. The imaginary 
values of $k_l$, denoted by $\kappa_l$ are to the right of the vertical 
axis and the imaginary values of $k_t$, denoted by $\kappa_t$ are 
below the horizontal axis in both figures. As in the rest of the paper, 
$d$ is the thickness of the plate. 
For these plots we used $c_l/c_t=1.66$, corresponding to SiN$_x$ plates. 
The $+$ and $-$ signs in the upper-right quadrants are only for orientation 
purposes.}
\label{klktfig}
\end{center}
\end{figure}

For symmetric Lamb modes, from Eq. (\ref{eqn_antisymmetric}) we 
calculate $k_l$ and $k_t$ as functions of $k_\parallel$. The results 
are shown in Fig. \ref{klktfig} (a). As expected, for each value of 
$k_\parallel$, the wave-vectors $k_l$ and $k_t$ take only discrete 
values, but not as simple as the values corresponding to the $h$ modes 
(\ref{vxsh}). Each curve in 
Fig. \ref{klktfig} (a) corresponds to a diffferent branch of the dispersion 
relation, $\omega_{k_\parallel,k_t(k_\parallel,m),s}$, 
where $m=0,1,\ldots$ denotes the branch number. Branches with bigger $m$ 
are placed above branches with smaller $m$.

As $k_\parallel$ increases, $k_l(k_\parallel,m)$ decreases and, after 
reaching the value 0, turns imaginary. On the other hand, 
$k_t(k_\parallel,m)$ first increases with 
$k_\parallel$, reaches a maximal value and then decreases monotonically 
as $k_\parallel$ increases to infinity. Its decrease is bounded for all 
branches, except the lowest one, where, after reaching zero at some finite 
$k_\parallel$, it turns imaginary in the lower-left quadrant of Fig. 
\ref{klktfig} (a). For the clarity of the calculations, the imaginary 
values of $k_l$ and $k_t$ are denoted as $i\kappa_l$ and $i\kappa_t$, 
respectively, where $\kappa_l$ and $\kappa_t$ take real, positive values. 
Both, $\kappa_l$ and $\kappa_t$ increase without limit as $k_\parallel$ 
increases to infinity. 

We encounter a similar situation for the antisymmetric Lamb modes (see 
Fig. \ref{klktfig} b). The only marked differences between the symmetric and 
antisymmetric modes are the following: 
the asymptotic values of $(d/2)\cdot k_t(k_\parallel\to\infty,m)$ are 
$\pi,2\pi,\ldots$ for symmetric modes and $\pi/2,3\pi/2,\ldots$ for 
antisymmetric modes; the maxima of $(d/2)\cdot k_t(k_\parallel,m)$ are 
$(2m+1)\pi/2$ for the symmetric modes and $m\pi$ for the antisymmetric modes. 
Also, for the antisymmetric modes $k_t(k_\parallel,0)$ (i.e. the lowest 
branch) takes only imaginary values. 

Integrating $|\bu_s|^2$ over the volume of the plate and using the 
transcendental equation (\ref{eqn_symmetric}), together with
Snell's law, 
$\omega^2=c_t^2(k_t^2+k_\parallel^2)=c_l^2(k_l^2+k_\parallel^2)$, 
we obtain the normalization constant for 
the symmetric modes in the \textit{quadrant I} (upper-right) of Fig. 
\ref{klktfig} (a) 
\begin{subequations} \label{Nsset}
\begin{eqnarray}
(N_s^I)^{-2} 
 &=& A\left\{
     4k_t^2k_\parallel^2\cos^2(k_td/2)\left(
     (k_l^2+k_\parallel^2)\frac{d}{2}
    -(k_l^2-k_\parallel^2)\frac{\sin(k_ld)}{2k_l}\right)\right.\nonumber\\
  && +(k_t^2-k_\parallel^2)^2\cos^2(k_ld/2)\left(
     (k_t^2+k_\parallel^2)\frac{d}{2}
    +(k_t^2-k_\parallel^2)\frac{\sin(k_td)}{2k_t}\right)\nonumber\\
  &&+4k_tk_\parallel^2(k_t^2-k_\parallel^2)\cos^2(k_ld/2)\sin(k_td)
    \left.\vphantom{\frac{1}{2}}\right\}\label{NsI} 
\end{eqnarray}
in the \textit{quadrant II} (upper-left), 
\begin{eqnarray}
(N_s^{II})^{-2}
 &=& A\left\{
     4k_t^2k_\parallel^2\cos^2(k_td/2)\left(
     (\kappa_l^2+k_\parallel^2)\frac{\sinh(\kappa_ld)}{2\kappa_l}
    -(\kappa_l^2-k_\parallel^2)\frac{d}{2}\right)\right.\nonumber\\
  && +(k_t^2-k_\parallel^2)^2\cosh^2(\kappa_ld/2)\left(
     (k_t^2+k_\parallel^2)\frac{d}{2}
    +(k_t^2-k_\parallel^2)\frac{\sin(k_td)}{2k_t}\right)\nonumber\\
  &&+4k_tk_\parallel^2(k_t^2-k_\parallel^2)\cosh^2(\kappa_ld/2)\sin(k_td)
    \left.\vphantom{\frac{1}{2}}\right\}\label{NsII} 
\end{eqnarray}
and in the \textit{quadrant III} (lower-left), 
\begin{eqnarray}
(N_s^{III})^{-2}
 &=& A\left\{
     4\kappa_t^2k_\parallel^2\cosh^2(\kappa_td/2)\left(
     (\kappa_l^2+k_\parallel^2)\frac{\sinh(\kappa_ld)}{2\kappa_l}
    -(\kappa_l^2-k_\parallel^2)\frac{d}{2}\right)\right.\nonumber\\
  && +(\kappa_t^2+k_\parallel^2)^2\cosh^2(\kappa_ld/2)\left(
     (\kappa_t^2+k_\parallel^2)\frac{\sinh(\kappa_td)}{2\kappa_t}
    +(\kappa_t^2-k_\parallel^2)\frac{d}{2}\right)\nonumber\\
  &&-4\kappa_tk_\parallel^2(\kappa_t^2+k_\parallel^2)\cosh^2(\kappa_ld/2)
     \sinh(\kappa_td)
    \left.\vphantom{\frac{1}{2}}\right\}\label{NsIII} 
\end{eqnarray}
\end{subequations}

Similarly, the normalization constants for symmetric modes in the 
three quadrants are 
\begin{subequations} \label{Naset}
\begin{eqnarray}
(N_a^{I})^{-2}
 &=& A\left\{
     4k_t^2k_\parallel^2\sin^2(k_td/2)\left(
     (k_l^2+k_\parallel^2)\frac{d}{2}
    +(k_l^2-k_\parallel^2)\frac{\sin(k_ld)}{2k_l}\right)\right.\nonumber\\
  && +(k_t^2-k_\parallel^2)^2\sin^2(k_ld/2)\left(
     (k_t^2+k_\parallel^2)\frac{d}{2}
    -(k_t^2-k_\parallel^2)\frac{\sin(k_td)}{2k_t}\right)\nonumber\\
  && -4k_tk_\parallel^2(k_t^2-k_\parallel^2)\sin^2(k_ld/2)\sin(k_td)
     \left.\vphantom{\frac{1}{2}}\right\}\label{NaI} \\
(N_a^{II})^{-2}
 &=& A\left\{
     4k_t^2k_\parallel^2\sin^2(k_td/2)\left(
     (\kappa_l^2+k_\parallel^2)\frac{\sinh(\kappa_ld)}{2\kappa_l}
    +(\kappa_l^2-k_\parallel^2)\frac{d}{2}\right)\right.\nonumber\\
  && +(k_t^2-k_\parallel^2)^2\sinh^2(\kappa_ld/2)\left(
     (k_t^2+k_\parallel^2)\frac{d}{2}
    -(k_t^2-k_\parallel^2)\frac{\sin(k_td)}{2k_t}\right)\nonumber\\
  && -4k_tk_\parallel^2(k_t^2-k_\parallel^2)\sinh^2(\kappa_ld/2)\sin(k_td)
     \left.\vphantom{\frac{1}{2}}\right\}\label{NaII} 
\end{eqnarray}
and 
\begin{eqnarray}
(N_a^{III})^{-2}
 &=& A \left\{
     4\kappa_t^2k_\parallel^2\sinh^2(\kappa_td/2)\left(
     (\kappa_l^2+k_\parallel^2)\frac{\sinh(\kappa_ld)}{2\kappa_l}
    +(\kappa_l^2-k_\parallel^2)\frac{d}{2}\right)\right.\nonumber\\
  && +(\kappa_t^2+k_\parallel^2)^2\sinh^2(\kappa_ld/2)\left(
     (\kappa_t^2+k_\parallel^2)\frac{\sinh(\kappa_td)}{2\kappa_t}
    -(\kappa_t^2-k_\parallel^2)\frac{d}{2}\right)\nonumber\\
  && -4\kappa_tk_\parallel^2(\kappa_t^2+k_\parallel^2)\sinh^2(\kappa_ld/2)
     \sinh(\kappa_td)
     \left.\vphantom{\frac{1}{2}}\right\}\label{NaIII} 
\end{eqnarray}
\end{subequations}

As one can notice, in the quadrants $II$ and $III$ the trigonometric 
functions are replaced by hyperbolic functions because of the change 
$k_l\to i\kappa_l$--in quadrant $II$--and $k_l\to i\kappa_l$, 
$k_t\to i\kappa_t$--in quadrant $III$. 

Although Eqs.\ (\ref{Nsset}) and (\ref{Naset}), with all their cases, 
are convenient because they are ready to use in practical calculations, 
we shall pack each set in one equation with the help of the complex 
wave-vector components 
$\bar{k}_t\equiv k_t+i\kappa_t$ and 
$\bar{k}_l\equiv k_l+i\kappa_l$: 
\begin{subequations}\label{eqn_N_complex}
\begin{eqnarray}
N_s^{-2}
 &=& A \left\{
     4|\bar{k}_t|^2k_\parallel^2|\cos(\bar{k}_td/2)|^2\left(
     (|\bar{k}_l|^2+k_\parallel^2)\frac{\sinh(\kappa_ld)}{2\kappa_l}
    -(|\bar{k}_l|^2-k_\parallel^2)\frac{\sin(k_ld)}{2k_l}\right)\right.\nonumber\\
  && +|\bar{k}_t^2-k_\parallel^2|^2|\cos(\bar{k}_ld/2)|^2\left(
     (|\bar{k}_t|^2+k_\parallel^2)\frac{\sinh(\kappa_td)}{2\kappa_t}
    +(|\bar{k}_t|^2-k_\parallel^2)\frac{\sin(k_td)}{2k_t}\right)\nonumber\\
  &&-4k_\parallel^2|\cos(\bar{k}_ld/2)|^2\left(\kappa_t(|\bar{k}_t|^2+k_\parallel^2)
     \sinh(\kappa_td)-k_t(|\bar{k}_t|^2-k_\parallel^2)\sin(k_td)\right)
    \left.\vphantom{\frac{1}{2}}\right\}\nonumber\\
  &&\label{eqn_N_s_complex}\\
N_a^{-2}
 &=& A \left\{
     4|\bar{k}_t|^2k_\parallel^2|\sin(\bar{k}_td/2)|^2\left(
     (|\bar{k}_l|^2+k_\parallel^2)\frac{\sinh(\kappa_ld)}{2\kappa_l}
    +(|\bar{k}_l|^2-k_\parallel^2)\frac{\sin(k_ld)}{2k_l}\right)\right.\nonumber\\
  && +|\bar{k}_t^2-k_\parallel^2|^2|\sin(\bar{k}_ld/2)|^2\left(
     (|\bar{k}_t|^2+k_\parallel^2)\frac{\sinh(\kappa_td)}{2\kappa_t}
    -(|\bar{k}_t|^2-k_\parallel^2)\frac{\sin(k_td)}{2k_t}\right)\nonumber\\
  && -4k_\parallel^2|\sin(\bar{k}_ld/2)|^2\left(\kappa_t(|\bar{k}_t|^2+k_\parallel^2)
     \sinh(\kappa_td)+k_t(|\bar{k}_t|^2-k_\parallel^2)\sin(k_td)\right)
     \left.\vphantom{\frac{1}{2}}\right\} \nonumber \\
&& \label{eqn_N_a_complex}\,.
\end{eqnarray}
\end{subequations}
Equations (\ref{eqn_N_complex}) are very general. For the phonon modes
(i.e. for real $k_\parallel$), $\bar{k}_t$ and $\bar{k}_l$ are either 
real or imaginary and the normalization constants (\ref{Nsset}) and 
(\ref{Naset}) can be extracted from (\ref{eqn_N_complex})
by \textit{taking the limit} of the redundant component of $\bar{k}_t$ and 
$\bar{k}_l$ going to zero.
The normalization constants (\ref{Nshset}), (\ref{Nsset}), and 
(\ref{Naset}) are chosen so that 
$||\bu_{\bk_\parallel,k_t,\sigma}||=1$, for any 
$\sigma,\bk_\parallel$ and $k_t$. In the next section 
the wave-functions will be multiplied by still another 
constant, which will give the right dimensions to the phonon field. 

\section{Quantization of the elastic field} \label{quantization}

For the quantization of the elastic field we start from the classical 
Hamiltonian: 
\begin{equation}
U = \int_V\left(\frac{\rho \dot u^2}{2}+\frac{S_{ij} c_{ijkl}
S_{kl}}{2} \right) \,. \label{Uel}
\end{equation}
where $S_{ij}$ are the components of the strain field, which is 
the \textit{symmetric gradient} of the displacement field, 
$S_{ij}\equiv(\nabla_S\bu)_{ij}=(\partial_i u_j+\partial_j u_i)/2$. 
The canonical variables are the field, $\bu$, and the conjugate 
momentum, $\bpi=\rho\bu$, which satisfy the Hamilton equations 
\begin{subequations} \label{Hamiltonset}
\begin{eqnarray}
\dot\bu &=& \frac{\delta U}{\delta\bpi} \,,\label{Hamilton1} \\
\dot{\bpi} &=& -\frac{\delta U}{\delta\bu} \,. \label{Hamilton2}
\end{eqnarray}
\end{subequations}
Equation (\ref{Hamilton2}) is nothing but the dynamic equation (\ref{wave1}). 

In the second quantization, $\bu$ and $\bpi$ become the field operators, 
$\tilde\bu$ and $\tilde{\bpi}$, respectively. If we denote by 
$\tilde b^\dagger_{\bk_\parallel,k_t,\sigma}$ and 
$\tilde b_{\bk_\parallel,k_t,\sigma}$, the creation and 
annihilation operators 
of a phonon with quantum numbers $\bk_\parallel,\ k_t$ and 
polarization $\sigma$ (in the notation that we used before), 
then the \textit{real} displacement and generalized momentum field 
operators, $\tilde\bu(\br)=\tilde\bu^\dagger(\br)$ and 
$\tilde{\bpi}(\br)=\tilde{\bpi}^\dagger(\br)$, are 
\begin{subequations} \label{fildop}
\begin{equation}
\tilde\bu(\br) = \sum_{\bk_\parallel,k_t,\sigma}\left[
\ff_{\bk_\parallel,k_t,\sigma}(\br) 
\tilde b_{\bk_\parallel,k_t,\sigma} + 
\ff^*_{\bk_\parallel,k_t,\sigma}(\br) 
\tilde b^\dagger_{\bk_\parallel,k_t,\sigma} \right]
 \label{displacementop}
\end{equation}
and 
\begin{eqnarray}
\tilde{\bpi}(\br) &=& \rho\sum_{\bk_\parallel,k_t,\sigma} \left[
\dot\ff_{\bk_\parallel,k_t,\sigma}(\br) 
\tilde b_{\bk_\parallel,k_t,\sigma} + 
\dot\ff^*_{\bk_\parallel,k_t,\sigma}(\br) 
\tilde b^\dagger_{\bk_\parallel,k_t,\sigma} \right] \nonumber \\
&=& -i\rho\sum_{\bk_\parallel,k_t,\sigma}
\omega_{\bk_\parallel,k_t,\sigma} \left[ 
\ff_{\bk_\parallel,k_t,\sigma}(\br) 
\tilde b_{\bk_\parallel,k_t,\sigma} -
\ff_{\bk_\parallel,k_t,\sigma}(\br) 
\tilde b_{\bk_\parallel,k_t,\sigma} \right] \,, 
\label{momentumop}
\end{eqnarray}
\end{subequations}
where $\ff_{\bk_\parallel,k_t,\sigma}(\br)\equiv C\bu_{\bk_\parallel,k_t,\sigma}(\br)$ and $C$ is a real constant which we shall 
determine from the commutation relations of the $\tilde b$ operators. 
In Eqs.\ (\ref{fildop}) we do not take $k_l$ as a summation variable, since 
this is either zero (for $h$ fields), or is determined by 
$\bk_\parallel$ and $k_t$ (via Eq.\ (\ref{eqn_symmetric}) or 
Eq.\ (\ref{eqn_antisymmetric}) for the symmetric and antisymmetric case, 
respectively). 

From Eqs.\ (\ref{fildop}) we can extract the operators $\tilde b$ and 
$\tilde b^\dagger$ in terms of $\tilde \bu$ and $\tilde {\bpi}$. In order to do this, let's first note from (\ref{eigenmodes}) that 
\begin{subequations}
\begin{eqnarray}
[\bu_{\bk_\parallel,k_t,h}(\br)]^*
&=&\bu_{-\bk_\parallel,k_t,h}(\br), \label{complexush} 
\end{eqnarray}
whereas 
\begin{eqnarray}
[\bu_{\bk_\parallel,k_t,s}(\br)]^* 
=\bu_{-\bk_\parallel,k_t,s}(\br) 
&{\rm or}& 
[\bu_{\bk_\parallel,k_t,s}(\br)]^* 
=-\bu_{-\bk_\parallel,k_t,s}(\br) , \label{complexus} 
\end{eqnarray}
depending weather $k_t$ is real or imaginary, respectively, and 
\begin{eqnarray}
[\bu_{\bk_\parallel,k_t,a}(\br)]^* 
=\bu_{-\bk_\parallel,k_t,a}(\br) 
&{\rm or}& 
[\bu_{\bk_\parallel,k_t,a}(\br)]^* 
=-\bu_{-\bk_\parallel,k_t,a}(\br) , \label{complexua} 
\end{eqnarray}
\end{subequations}
depending weather $k_l$ is real or imaginary, respectively. So 
\begin{equation}
\int_V \ff^T_{\bk_\parallel,k_t,\sigma}(\br)\cdot 
\ff_{\bk'_\parallel,k_t',\sigma'}(\br)\,d^3\br = 
a_{\sigma,k_t,k_l}C^2 \delta_{\sigma,\sigma'} 
\delta_{\bk_\parallel,-\bk'_\parallel}
\delta_{k_t,k_t'}
\end{equation}
where by $\ff^T$ we denote the transpose of the vector $\ff$ and 
$a_{\sigma,k_t,k_l}=\pm1$, according to Eqs.\ (\ref{complexus}) and 
(\ref{complexua}). 
Multiplying (\ref{displacementop}) and (\ref{momentumop}) by 
$\ff^\dagger_{\bk_\parallel,k_t,\sigma}(\br)$  and integrating 
over $V$, we get 
\begin{subequations} \label{setb}
\begin{eqnarray}
\int_V \ff^\dagger_{\bk_\parallel,k_t,\sigma}(\br)\tilde\bu(\br)
\,d^3\br &=& C^2 [\tilde b_{\bk_\parallel,k_t,\sigma} + 
a_{\sigma,k_t,k_l} \tilde 
b^\dagger_{-\bk_\parallel,k_t,\sigma}] \,,\label{intVbu} \\
\int_V \ff^\dagger_{\bk_\parallel,k_t,\sigma}(\br)\tilde{\bpi}(\br)
\,d^3\br &=& -i\rho\omega_{\bk_\parallel,k_t,\sigma}C^2\left[
\tilde b_{\bk_\parallel,k_t,\sigma} - 
a_{\sigma,k_t,k_l} \tilde b^\dagger_{-\bk_\parallel,k_t,\sigma} 
\right]\,.\label{intVbpi} 
\end{eqnarray}
\end{subequations}
Solving the system we obtain  
\begin{equation}
\tilde b_{\bk_\parallel,k_t,\sigma} = \frac{1}{2C^2}\left[
\int_V \ff^\dagger_{\bk_\parallel,k_t,\sigma}(\br)\tilde\bu(\br)
\,d^3\br + \frac{i}{\rho\omega_{\bk_\parallel,k_t,\sigma}}
\int_V \ff^\dagger_{\bk_\parallel,k_t,\sigma}(\br)\tilde{\bpi}(\br)
\,d^3\br \right]
\end{equation}
and obviously, 
\begin{equation}
\tilde b^\dagger_{\bk_\parallel,k_t,\sigma} = \frac{1}{2C^2}\left[
\int_V \ff^T_{\bk_\parallel,k_t,\sigma}(\br)\tilde\bu(\br)
\,d^3\br - \frac{i}{\rho\omega_{\bk_\parallel,k_t,\sigma}}
\int_V \ff^T_{\bk_\parallel,k_t,\sigma}(\br)\tilde{\bpi}(\br)
\,d^3\br \right] \,.
\end{equation}
Using the canonical commutation relations, 
$[\tilde\bu(\br),\tilde\bu(\br')]=[\tilde{\bpi}(\br),\tilde{\bpi}(\br')]=0$ 
and $[\tilde\bu(\br),\tilde{\bpi}(\br')]=i\hbar\delta(\br-\br')$, 
we obtain the commutation relations for the operators $\tilde b$ and 
$\tilde b^\dagger$: 
\[
[\tilde b_{\bk_\parallel,k_t,\sigma},
\tilde b_{\bk'_\parallel,k_t',\sigma'}] = 
[\tilde b^\dagger_{\bk_\parallel,k_t,\sigma},
\tilde b^\dagger_{\bk'_\parallel,k_t',\sigma'}] = 0 
\]
and 
\[
[\tilde b_{\bk_\parallel,k_t,\sigma},
\tilde b^\dagger_{\bk'_\parallel,k_t',\sigma'}] = 
\delta_{\sigma,\sigma'}\delta_{\bk_\parallel,\bk'_\parallel}
\delta_{k_t,k_t'}, 
\]
provided that 
\begin{equation}
C = \sqrt{\frac{\hbar}{2\rho\omega_{\bk_\parallel,k_t,\sigma}}} \,.
\label{constantvalue}
\end{equation}

Using Eqs.\ (\ref{fildop}), with the proper normalization of $\ff$, 
we can write $U$ (\ref{Uel}) in operator form, 
\begin{eqnarray}
U &=& \frac{\rho}{2}\int_V d^3\br\sum_{\bk_\parallel,k_t,\sigma}\sum_{\bk'_\parallel,k_t',\sigma'} 
\left[\dot\ff^\dagger_{\bk_\parallel,k_t,\sigma}(\br) \tilde 
b^\dagger_{\bk_\parallel,k_t,\sigma} + 
\dot\ff^T_{\bk_\parallel,k_t,\sigma}(\br) 
\tilde b_{\bk_\parallel,k_t,\sigma}
\right]\nonumber \\
&& \times \left[\dot\ff_{\bk'_\parallel,k_t',k_l',\sigma'}(\br) 
\tilde b_{\bk'_\parallel,k_t',k_l',\sigma'} +
\dot\ff^\star_{\bk'_\parallel,k_t',k_l',\sigma'}(\br) 
\tilde b^\dagger_{\bk'_\parallel,k_t',k_l',\sigma'}\right] 
\nonumber \\
&&+ \frac{1}{2}\int_V d^3\br\sum_{\bk_\parallel,k_t,\sigma} 
\sum_{\bk'_\parallel,k_t',\sigma'}\left[\partial_i 
[\ff^\star_{\bk_\parallel,k_t,\sigma}(\br)]_j 
\tilde b^\dagger_{\bk_\parallel,k_t,\sigma} + 
\partial_i [\ff_{\bk_\parallel,k_t,\sigma}(\br)]_j 
\tilde b_{\bk_\parallel,k_t,\sigma}\right]\nonumber \\
&&\times c_{ijkl}\cdot\left[
\partial_k [\ff_{\bk_\parallel,k_t,\sigma}(\br)]_l 
\tilde b_{\bk_\parallel,k_t,\sigma} + 
\partial_k [\ff^\star_{\bk_\parallel,k_t,\sigma}(\br)]_l 
\tilde b^\dagger_{\bk_\parallel,k_t,\sigma}\right] \nonumber \\
&=& \sum_{\bk_\parallel,k_t,\sigma}\hbar
\omega_{\bk_\parallel,k_t,\sigma}
[\tilde b^\dagger_{\bk_\parallel,k_t,\sigma}
\tilde b_{\bk_\parallel,k_t,\sigma}+1/2] \,.
\label{diagham}
\end{eqnarray}
As expected, the Hamiltonian of the elastic body can be written as a sum 
of Hamiltonians of harmonic oscillators. These oscillators are the phonon 
modes of the plate. 

We use this formalism elsewhere to describe the interaction of phonons 
with the disorder in amorphous materials 
\cite{submitted.bulkTLS,submitted.TLS_Lamb}. 

\section{Conclusions} \label{conclusions}

The vibrational modes of a thin plate (\ref{eigenmodes}) are well known 
from elasticity theory \cite{Auld:book}. The purpose of the paper is 
to quantize the elastic field and for this we have to know if these modes, 
or part of them, form a complete set of orthogonal functions. But since 
the modes are the solutions 
of the eigenvalue-eigenvector problem of the operator $\tilde L$ 
(\ref{wave2}), we showed that they form a complete set by proving that 
$\tilde L$ is self-adjoint. 

Nevertheless, not all the functions of the form (\ref{eigenmodes}) 
are orthogonal to each-other, so to build the complete, 
orthogonal set of functions, we made use of a generic ``momentum'' 
operator, $\tilde\bk_\parallel\equiv i(\partial_x+\partial_y)$, which 
commutes with $\tilde L$. Since for a plate with infinite lateral extension 
or a finite rectangular plate with periodic boundary conditions at the edges 
the operator $\tilde\bk_\parallel$ is also self-adjoint, $\tilde L$ and 
$\tilde\bk_\parallel$ admit a common, complete set of orthogonal 
eigenfunctions. The degenerate eigenvalues of $\tilde\bk_\parallel$ 
are, of course, the wave-vectors parallel to the plate surfaces, 
$\bk_\parallel$, of real components. Therefore, the complete set of 
eigenfunctions are the ones given by Eqs.\ (\ref{eigenmodes}), with 
real $\bk_\parallel$. 

In Section \ref{orthogonality}, based on the hermiticity of the operator 
$\tilde L$, we showed that these functions (\ref{eigenmodes}) are indeed 
orthogonal to each-other and in Section \ref{normalization} we 
calculated the normalization factors. 

Having all these ingredients, in Section \ref{quantization} we 
presented the formal quantization procedure which is applied 
elsewhere \cite{submitted.bulkTLS,submitted.TLS_Lamb} to calculate the thermal 
properties of ultra-thin plates at low temperatures, and to deduce
some of the observed features of the 
\textit{standard tunneling model} in bulk amorphous materials.

\section{Acknowledgements}

We are grateful to Profs. N. Angelescu, Y. M. Galperin and M. Manninen 
for helpful discussions. DVA acknowledges the NATO support, 
by the grant EAP.RIG 982080.


\end{document}